\documentclass[sigconf,natbib=true]{acmart} %

\usepackage[utf8]{inputenc} 
\usepackage[T1]{fontenc}    
\usepackage{hyperref}       
\usepackage{url}
\usepackage{multirow}
\usepackage{booktabs}       
\usepackage{amsfonts}       
\usepackage{nicefrac}       
\usepackage{microtype}      
\usepackage{xcolor}         
\usepackage{tabularx}
\usepackage{notes-alt} 
\usepackage{multirow}
\usepackage{amsmath}
\usepackage{graphicx}
\graphicspath{ {imgs/} }
\usepackage{svg}
\usepackage{enumitem}
\usepackage{subcaption}
\usepackage{amsthm, amsmath,algorithm2e}
\usepackage{xcolor}

\definecolor{blue}{RGB}{17,220,247}
\definecolor{purple}{RGB}{163,115,250}

\usepackage{wrapfig}
\newcommand{\mpara}[1]{\medskip\noindent{\bf #1}}

\newcommand{\bart}{\textsc{BART}}
\newcommand{\gpt}{\textsc{GPT2}}
\newcommand{\gptt}{\textsc{GPT3}}

\newcommand{\mi}{\textsf{ machine intent}}

\AtBeginDocument{%
  \providecommand\BibTeX{{%
    \normalfont B\kern-0.5em{\scshape i\kern-0.25em b}\kern-0.8em\TeX}}}
    
\copyrightyear{2023}
\acmYear{2023}
\acmConference[Gen-IR@SIGIR2023]{The First Workshop on Generative Information Retrieval}{July 27, 2023}{Taipei, Taiwan}
\acmBooktitle{The First Workshop on Generative Information Retrieval (Gen-IR@SIGIR23), July 27, 2023, Taipei, Taiwan}
\acmDOI{}
\acmPrice{}
\acmISBN{}
\setcopyright{rightsretained}

\title{Query Understanding in the Age of Large Language Models}

\author{Avishek Anand}
\email{Avishek.Anand@tudelft.nl}
\affiliation{%
  \institution{Delft University of Technology}
  \country{Netherlands}
}

\author{Venktesh V}
\email{v.Viswanathan-1@tudelft.nl}
\affiliation{%
  \institution{Delft University of Technology}
  \country{Netherlands}
}

\author{Abhijit Anand}
\email{aanand@l3s.de}
\affiliation{%
  \institution{L3S Research Institute}
  \country{Germany}
}
\author{Vinay Setty}
\email{vsetty@acm.org}
\affiliation{%
  \institution{University of Stavanger}
  \country{Norway}
}

\begin{document}

\begin{abstract}

Querying, conversing, and controlling search and information-seeking interfaces using natural language are fast becoming ubiquitous with the rise and adoption of large-language models (LLM). 
In this position paper, we describe a generic framework for interactive query-rewriting using LLMs.
Our proposal aims to unfold new opportunities for improved and transparent intent understanding while building high-performance retrieval systems using LLMs.
A key aspect of our framework is the ability of the rewriter to fully specify the \textit{machine intent} by the search engine in natural language that can be further refined, controlled, and edited before the final retrieval phase.
The ability to present, interact, and reason over the underlying machine intent in natural language has profound implications on transparency, ranking performance, and a departure from the traditional way in which supervised signals were collected for understanding intents.
We detail the concept, backed by initial experiments, along with open questions for this interactive query understanding framework.

\end{abstract}
\maketitle

\section{Introduction}
\label{sec:intro}

The emergence of large language models (LLMs) such as GPT-3~\cite{brown2020gpt3} and GPT-4~\cite{openai2023gpt4} by OpenAI, PaLM~\cite{chowdhery:2022:palm} by Google and LLaMA~\cite{touvron2023llama} by Meta, has brought about a significant transformation in both academia and industry. 
These models have been fine-tuned to follow instructions with human feedback  and, consequently, have shown to significantly enhance understanding and generation capabilities for various language tasks~\cite{gpt3_prompting,gpt3incontext,ouyang2022training}. 
LLMs have also been adapted to present web search results interactively via applications such as Microsoft's Bing chat\footnote{\url{https://blogs.bing.com/search/march_2023/Confirmed-the-new-Bing-runs-on-OpenAI’s-GPT-4}} and Google's Bard\footnote{\url{https://bard.google.com}}. 

Leveraging LLMs for enhancing various components of web search systems has emerged as a promising area of research~\cite{metzler2021:rethinking,khattab2022:dsp,he2022rethink:retrieve,bonifacio2022inpars,dai2022promptagator,mackie2023generative}. For example, \citet{metzler2021:rethinking} introduce a method to represent knowledge within a corpus using a model, thereby eliminating the need for traditional indexes and thereby transforming understanding capabilities for several IR tasks.  Similarly, proposals such as InPars~\cite{bonifacio2022inpars} and PromptaGator~\cite{dai2022promptagator} employ LLMs to generate paraphrased text, serving as data augmentation strategies that enhance the training of ranking models.

In this paper, we specifically look into rethinking query understanding component of the search systems by leveraging LLMs. 
Query understanding is an important component of search systems designed to bridge the gap between users' query intent and machine's query understanding. 

\begin{figure*}
    \centering
    \includegraphics[width=0.8\textwidth]{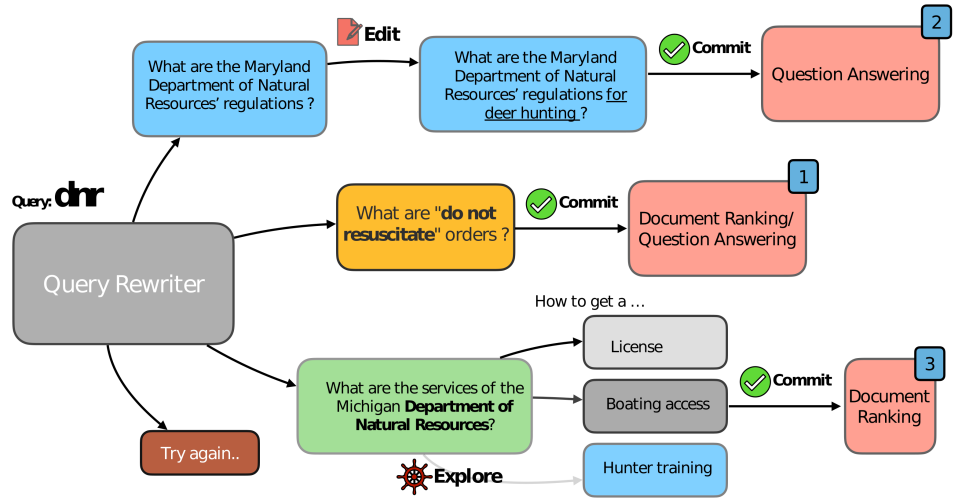}
    \caption{Different pathways to ranking or answering from an initial TREC Web Track query ``dnr'' --  1) Commit -- machine intent is consistent with the user intent, 2) Edit - when user can update the machine intent in place, 3) Explore/Refine: when user wants to refine the machine intent.}
    \label{fig:qr-edit}
\end{figure*}

The central problem of IR systems, also referred to as the ``holy grail'' of IR, is to overcome the \textit{vocabulary mismatch} between the user and the system \cite{wang2020deep}. 
This leads to the challenge of matching the user's information need needs with the relevant documents in the collection.
We aim to bridge this gap by proposing an interactive query rewriting framework using LLMs (shown in Figure \ref{fig:qr-edit}). 
 We define research directions that handle multiple challenges such as like minimizing query reformulations while allowing for low-cost user exploration.
 We believe that our vision of greatly improving interactive query understanding using LLMs has far-reaching implications for -- \textit{how users interact with search engines}, \textit{how data is collected from user interactions}, \textit{the quality of data and feedback}, and \textit{its impact on the learning ecosystem}.


In conventional search engines, the query understanding component comprises several subcomponents, including spell checkers, query classifiers, query expansions, and query suggestions~\cite{chang2020query,wang2020deep:query:rewrites}. Presently, users experience limited transparency and minimal control during the query understanding process. In most cases, the users interact with query suggestions and do not get to see the final rewritten query. This limits the users' ability to modify or reformulate the query to align with their own latent intent.

Moreover, current ranking systems offer indirect and delayed feedback. Users primarily provide feedback by clicking on a ranked list of documents, but clicks are subject to biases like position and trust biases~\cite{www2021-zheng:trust-bias,sigir2016-wang:exposure-bias,wsdm2017-joachims:position-bias}. Consequently, they are unreliable indicators of the user's evaluation of their information need, often leading to query reformulations after time-consuming document inspection. 

This position paper introduces an \textit{LLM-based interactive query understanding framework} that replaces traditional query rewriter components with a single LLM-based component. This approach minimizes the potential for error propagation arising from multiple components in the system. We investigate the significant implications of using such an LLM-based query rewriter for designing search systems and its impact on their performance. 

Our conceptualized \textit{query rewriter} outputs machine intents-- or the machine's interpretation of the user intent-- that are fully specified in natural language and describes the machine intent accurately. 
The output intents of our query rewriter allow for natural-language and editable explanations, facilitating easy user feedback.

Our main contributions are outlined below:
\begin{enumerate}
    \item A proposal to replace the multi-component query understanding pipeline with a monolithic LLM that can perform versatile query transformations. We explore research directions to minimize query reformulations and failed search missions while enabling low-cost user exploration.
    \item We strive to ensure that the \textit{machine intents are faithful, plausible, and controllable} in our proposal. In Section~\ref{sec:perspective-interpretability}, we discuss research directions for enforcing interpretability constraints within the context of explainable IR.
    \item We propose a design for a framework which incorporates direct and indirect feedback to fine-tune the rewriter and ranker to yield a mutually reinforcing relationship between the components. This helps reconcile user and machine intent yielding relevant documents.
\item A qualitative feasibility study demonstrating how LLMs can generate plausible rewrites of original queries which can potentially improve downstream retrieval performance.
\end{enumerate}

Our results show that even simple data augmentation, prompting, and in-context learning approaches (in Section \ref{experiments}) results in highly plausible machine intents with no impact on ranking performance.
We conclude the paper by detailing open problems and the challenges that need to be tackled, which open up exciting future research directions (Section \ref{open_questions}).




\section{Novel Query Rewriting Framework}
\label{sec:framework}
We propose a new paradigm for query understanding for keyword search and answer generation, where the aim of the re-written query is to present a plausible natural language descriptions of the underlying machine intent to the user.
These natural language machine intents serve the purpose to elicit early user feedback, refinement, and reformulation prior to the actual re-ranking or answer generation operation.
To start off, the query rewriter component (a generative model) generates multiple and possibly diverse machine intents of the given user query.
For example, Figure~\ref{fig:qr-edit} shows some of the possible actions that the user can take once shown the underlying intents.
In the best case, the user can \textbf{``commit''} to one of the machine intents, leading to a final ranking or answer generation phase. 
Alternately, three other actions are possible when the machine intent does not align with the user's actual query intent, as shown in Figure~\ref{fig:qr-edit}.

\subsection{Soft rewrites}
\label{sec:soft-rewrite}

If the machine intent is close to the user's intent but not specific enough to yield the intended response, the user can edit the rewritten query. We term this mode as a \textbf{``soft rewrite''} mode, where the user only deems minor edits as necessary to the machine generated rewrites. For instance, in Figure~\ref{fig:qr-edit}, in the first search scenario, the user finds the rewrite relevant as the intent is to retrieve documents related to \textit{department of natural resources (dnr)}. However, the user is looking for specifics of regulations regarding \textit{deer hunting}. Since the query only requires a minor edit, it is a soft rewrite. Hence, the user can customize the generated natural language summaries in our framework.

\subsection{Refinement and Exploration}
\label{sec:exploration}

While minor edits can aid in minor query customizations, but the intents can still be either non-specific or multi-faceted.
For instance, in Figure \ref{fig:qr-edit}, the intent in the third search scenario is too generic and the user, in fact, interested in one of the ``services of department of natural resources''. 
The user at this stage decides to \textbf{explore} and refine the current state of the machine intent. 
Note that further refinements are always conditioned on the current search state of the user, instead of being considered independently.
Eventually, the user can commit to a specific query rewrite that satisfies its information need. 

\subsection{Hard rewrites}
\label{sec:hard-rewrites}

However, there is always a possibility that none of the query rewrites are relevant to the user. In this case, the user may ignore the suggestions and reformulate the query altogether. 
This is termed as a \textbf{"hard rewrite"}. 
Note that hard re-writes are akin to query reformulations in current search systems but without the retrieval step. 
The aim of our query re-writing framework is to reduce the cost of failed searches by exposing the user to the machine intent as an early intervention step.
The user therefore does not have to inspect the ranked list and can potentially iterate faster based on the interactive query understanding step.

\subsection{A Generative Query Re-writer}
\label{design}

The core of our design decisions are based on implementing the query re-writer based on an LLM that we believe is a versatile generative model. 
A major outcome of our framework is that the query edits serve as early feedback that can be used to train, personalize and fine-tune the re-writer to the user preferences and needs. 
Various design considerations must be evaluated when training and customizing an LLM-based query re-writer, and this can have far-reaching implications on task-specific models downstream, such as the re-ranker and reader.  
We now list the main design choices:

\begin{description}[leftmargin=0.25cm]
\item[Pre-trained rewriter:] The rewriter could be large language models pre-trained on web-scale data and could be coupled with any off-the-shelf zero-shot large-language model~\cite{gpt3_prompting,web_gpt}. 
The rewrites could be generated through prompt engineering, in-context learning, or chain-of-thought reasoning~\cite{gpt3incontext}.
 \item[Fine-tuned rewriter:] The rewriter could be fine-tuned on the user search profile to characterize the queries and also implicitly cluster similar search profiles. The rewriter could then be employed to generate \textit{personalized rewrites} with little effort. The intents would reflect user choices, possibly alleviating the need for hard rewrites.
 \item[Retrieval augmented rewrites:] Instead of relying purely on parametric memory, the query understanding phase can also be augmented with a retrieval component~\cite{zamani2022:reml}.
Although generation from parametric memory produces plausible text, but a well acknowledged problem of LLMs is that of hallucination and lack of grounding~\cite{metzler2021:rethinking}. 
Towards this, an optional retrieval module \cite{retro} could be coupled with the interactive query understanding phase for more granular and relevant rewrites. 
\end{description}

\section{Impact on query modelling}
\label{sec:query-modelling}

The brevity of search queries and the potential ambiguity of user intent pose challenges for search systems to retrieve relevant documents. Search systems apply a wide variety of techniques to address this issue~\cite{chang2020query}. In this section, we delve into the query understanding and rewriting methods commonly used in retrieval tasks and conversational settings.

\vspace{-2.5mm}
\subsection{Query Understanding for Retrieval}

The goal of query rewriting is to align the query with the user intent. Both shallow and deep neural models are proposed for the query rewriting. Shallow models use substitution-based methods, which replace the query by substituting terms and phrases with synonyms and related phrases~\cite{jones2006generating}. A closely related method is query expansion, which adds relevant terms using the pseudo-relevance feedback~\cite{lv2009comparative}.  More modern approaches use word embeddings to rewrite queries~\cite{grbovic2015context}. There are also generative models which use Seq2Seq models~\cite{he2016learning} and reinforcement learning for generating rewritten queries~\cite{nogueira2017task}. Auto-completion and query suggestions are ways to guide the user to resolve the query intent more iteratively. There is a rich literature which suggests that these methods improve search performance~\cite{baeza2004query}. Unlike these methods, which rewrite the queries as keyword queries, our work focuses on generating natural language queries which are faithful and scrutable to the user intent.

Natural language question generation for query understanding is a relatively new trend in IR. \cite{rao2018learning} introduced a model to generate questions that clarify missing information in a specific setting. They used a reinforcement learning model to maximize a utility function based on the value added by the potential response to the question. \cite{trienes2019identifying}, on the other hand, focused on identifying unclear posts in a community question answering setting that require further clarification. More recently, \cite{arabzadeh2021matches} proposed to reformulate queries based on relevance judgements to provide performance gain by aligning with user intent. Controllable query reformulation is also of immense interest to promote content discovery. The authors of the work, \cite{content_retrievability}, propose a query generation approach to tackle retrievability bias by promoting generation of new entities in reformulated query. Query reformulation approaches have been of huge interest in the e-commerce domain and mostly require relevance feedback from the user for personalization \cite{query_expansion_e_commerce, taobao_search}. While these methods are domain-specific, \cite{zamani2020generating} propose using template-based slot-filling using both weakly supervised and reinforcement learning for open domain IR. Generating clarifying questions is more commonly used in conversational setting, which we discuss further in Section \ref{subsec:conv}. 

Although the methods mentioned above enhance the comprehension of queries, they cannot always substitute for one another, and combining them in sequential manner in search systems can introduce errors retrieving information. Our goal is to minimize the number of components in search systems for capturing the user query intent by natural language intent generation.

\subsection{Conversational Query Understanding}
\label{subsec:conv}

Conversational query rewriting is aimed at transforming a brief, context-dependent question into a fully specific, context-independent query that can be processed effectively by existing question answering and information retrieval systems. \cite{elgohary2019can}~introduced the task of conversational query rewriting with the CANRAD dataset, with the goal of making the queries self-contained. Several approaches exploit this dataset to improve the performance of conversational settings. \cite{vakulenko2021question}~propose a generative approach to replace tokens in the question to make it context-independent. Another approach uses few-shot fine-tuning of LLMs to generate queries that improve conversational search performance~\cite{yu2020few}. Others like ConQRR (Conversational Query Rewriting) rewrite a conversational question in context into a standalone question using a novel reward function to optimize retrieval models directly via reinforcement learning~\cite{wu2021conqrr}. More controlled generation approaches  have also been proposed \cite{Hao2022} to personalize rewrites to users. These methods mainly focus on resolving co-references of pronouns, which is complimentary to our approach in which we need to infer query intent from limited context.

Query suggestion is another way to guide users towards the right query in a conversational setting~\cite{rosset2020leading, sordoni2015hierarchical}. Asking clarifying questions is widely used for query understanding in conversational settings, which allows the users to provide additional context~\cite{bi2021asking, zamani2022conversational}. While these methods need training data resulting in cold start problems, \cite{wang2023zero} introduce zero-shot clarifying question generation.

Our approach does not rely on conversation context or prior answers, and instead works with under-specified queries to align human intent with machine intent.

\subsection{Large Language Generative Models}

Recent advances in neural text generation have demonstrated that large language models (LLMs) trained in a self-supervised generative fashion generalize well to downstream tasks in a few-shot or zero-shot setting \cite{brown2020language, wei2021finetuned, alayrac:2022:flamingo}. Prompting is shown to be an effective way for controlled text generation, zero-shot and few-shot learning of downstream tasks. In dense retrieval, Hyde \cite{hyde} demonstrates impressive zero-shot performance. The approach instructs a prompt based language model like \textsc{InstructGPT} to generate a hypothetical document followed by a dense retriever \cite{contriever} which introduces a bottleneck aiding to retrieve relevant documents. More recently, researchers have tested the effectiveness of LLMs for document expansion \cite{hyde, wang2023query2doc} in a zero-shot and few-shot setting for dense retrieval. However, these approaches have serious limitations at inference time in terms of efficiency due to autoregressive decoding of LLMs. These approaches are also prone to hallucination due to the generation of long context for ranking. Our approach focuses on disambiguating queries through fully specified natural language rewrites of the original queries. Our approach also requires no supervision in the form of relevance scores.

\subsection{Minimize Hard Query Rewrites}
We argue that query rewriting, whether in traditional IR or a conversational setting, provides limited flexibility to users. Despite the availability of several tools in existing search systems to capture query intent, users still need to rewrite their queries, resulting in hard rewrites as mentioned in Section \ref{sec:hard-rewrites}. Therefore, we propose the following research direction:

\mpara{RD I (Minimize Hard Rewrites):} \textit{How can we align machine intent with human intent while minimizing the need for hard query rewrites by the user?}

Current query rewriting methods spread tools across multiple components, making them less efficient for users. Our proposed solution is a single query rewriter that uses LLMs to create natural language query descriptions from vague keyword queries. These descriptions provide users with a clearer understanding of the machine's intent, enabling them to make more informed decisions and reducing the need for extensive hard query rewrites.

One of the challenges in RD I is to evaluate the effectiveness of the different modes of query rewriting mentioned in Section \ref{sec:feedback}. One way to assess scrutability is to use quantitative measures, such as evaluating how often the user commits to the query rewrite over soft rewriting and hard rewriting. 

\newcommand{\rewriter}{LLM}

\section{Impact on Explainable Rankers}
\label{sec:perspective-interpretability}

In the area of explainable IR, the central goal is to explain the reasoning behind the ranking decisions made by a ranker using an \textit{explanation} that is \textit{faithful} to the ranking model~\cite{anand2022:exir}.
Explanations are user-centric and can have multiple forms e.g., heatmaps or \textit{feature attributions}, extractive text snippets or \textit{rationales}, and term-based~\cite{singh:2020:fat:intentmodel,lyu:ecir:23:multiplex}. 
In some sense, most of EXIR approaches either directly~\cite{singh:2020:fat:intentmodel,lyu:ecir:23:multiplex} or indirectly~\cite{verma:2019:sigir:lirme,polley:2021:exdocs} try to uncover the often latent machine intent. 
Consequently, the natural language \mi{} from the \rewriter{} can naturally serve as an \textit{explanation} of the downstream ranking task if it is \textit{faithful} to the underlying ranking model.
The concept of faithfulness or fidelity is an important and necessary condition for explainers.
It refers to the ability of the explanation to truly capture the true behavior of the ranking model.
An explanation can be plausible or human understandable, even persuasive and convincing, but is not faithful to the ranking model if the ranker uses a different reasoning (different features, text patterns, shortcuts) for ranking. 

\subsection{Research Context}

Full interpretability while ensuring competitive performance is often hard, if not impossible, to attain with neural architectures. 
However, there has been progress in the area of explainable IR that are either \textit{interpretable-by-design} or \textit{posthoc methods}~\cite{zhang:2021:wsdm:expred,lei:2016:emnlp:rationalizing,leonhardt2021learnt}.
Posthoc methods for interpretability do not trade-off model performance for interpretability and are employed after model a ranking model is trained to posthoc analyse the reasons for the prediction. 
Unfortunately, most of the explanations of the posthoc methods are term-based and hence not fully specified for the end user.

More importantly, such methods are not entirely faithful to the trained ranking models. 
Moreover, it is also hard to evaluate if these models have faithfully deciphered the machine intent of the ranking models.

Models that are entirely faithful and provide a better specification of the query intent are explain-then-predict models. 
These models deliver partial interpretability using rationale-based explanations as an intermediate step. 
However, their explanations are not \textit{scrutable} and are limited to extractive pieces of text from documents in the ranked list. 

It is still the prerogative of the user to decipher the machine intent from the multiple (query, document) explanations. 
Our framework intends to build ranking models that are \textit{interpretable-by-design} where the machine intent can be fully specified in natural language for the user to scrutinize.

\begin{table}[ht]
    \centering
\begin{tabular}{lcccc}
\textbf{Methods} & \textbf{Plausible} & \textbf{Quality} & \textbf{Faithful} & \textbf{Editable}  \\
\hline

Feature attrib.~\cite{fernando:2019:sigir:studyshaponretrieval} & $\uparrow$ & $\uparrow$ & $\downarrow$ & - \\ 
Term-based~\cite{singh:2020:fat:intentmodel,singh:2019:wsdm:exs} & $\uparrow$ & $\uparrow$ & $\downarrow$ & - \\
Extractive ~\cite{zhang:2021:wsdm:expred,leonhardt:2021:arxiv:hard:kuma:ranking} & $\uparrow$ & $\downarrow$  & $\uparrow$ & - \\
\textbf{LLM-Rewriter}& $\uparrow$ & $\downarrow$  & $\uparrow$ & $\uparrow$ \\
\bottomrule
\end{tabular}
    \caption{Comparison of Interpretable models/approaches along the four dimensions desired.}
    \label{tab:datasets}
\end{table}




~\cite{radlinski2022natural} contrast the current approach of blackbox user models in recommender systems with a proposed scrutable user modeling allowing control of a system’s personalization.
Scrutable user models represent user models in natural language (NL) where scrutable language is defined as being both short and clear enough for someone to review and edit directly. Table ~\ref{tab:datasets} compares  different approaches based on \textbf{Plausible}, \textbf{Quality}, \textbf{Faithful} and \textbf{Editable} dimensions.

\begin{itemize}
    \item \textbf{Plausible or Human understandable:} Posthoc interpretability ~\cite{singh:2020:fat:intentmodel,fernando:2019:sigir:studyshaponretrieval,singh:2019:wsdm:exs}. High on understandability but low on plausibility.

    \item 
    \textbf{High performance:}
    posthoc methods are high on performance. Expred models trade-off performance and sparsity (plausibility)~\cite{leonhardt:2021:arxiv:hard:kuma:ranking,zhang:2021:wsdm:expred}. Expred models might not be high performance.

    \item
    \textbf{Faithful} Not only plausible. 
    Posthoc methods are plausible but not faithful. Expred models are fully faithful.

    \item
    \textbf{Scrutability.}
    Neither expred nor posthoc methods are scrutable. 
    CQA, e-cos, NLI datasets looked at generative models so they might be scrutable.
    
\end{itemize}

\subsection{Machine Intents as Explanations}

The primary question for realizing an interpretable machine intent is to make sure it is faithful to the model decision.
Towards this we define the following research direction:

\mpara{RD II (Faithful Machine Intents):} \textit{How do we ensure highly faithful \mi{} to the ranking model, while still being plausible and retaining good ranking performance.
}

A simple yet popular solution is to follow a \textit{rewrite-then-rank framework}  where a generative model is first trained to generate query explanations or query descriptions from an under-specified query. 
Subsequently, a neural ranker is trained to leverage the generated query explanation to retrieve relevant documents from the corpus.
To respect the faithfulness criterion, interpretability constraints have to be imposed over end-to-end training regimes.
Recently, proximal-policy optimization approaches~\cite{rlhf,ppo} have been proposed to train such chained models with arbitrary rewards. 
Rewards measuring the fidelity between intents and rankings can then be used to train both the rankers and the rewriters.
For example for the query ``Panther showtimes'' the generative LM generates ``Show the time slots for panther visit in local zoo''. But based on a scalar reward (human feedback) judging the relevance of returned documents the generator could rewrite the query intent to ``Show the time slots for black panther movie in theatre''.




\subsection{Rationale-based Models}

Rationale-based models in natural language processing (NLP) and IR are machine learning models designed to generate human understandable explanations or justifications for their predictions or decisions. 
Rationale-based models typically consist of two components: a rationale-generation model and a task model. 
Given the text input, the rationale-generation model generates an extractive or abstractive explanation. The task model subsequently uses just the generated explanation to make a prediction. The task model and the rationale generation module can be trained together or in a pipelined manner, depending on the specific application~\cite{zhang:2021:wsdm:expred}.
Rationalization in NLP was first introduced in 2007~\cite{zaidan2007using} with the objective of using annotator rationales to improve task performance for text categorization. 
Interestingly, explainability was not the core objective. However, explainability is an advantage of rationalization because it makes the model inherently explainable even if used in the context of task improvement.
~\cite{rajani2019explain} developed the Commonsense Auto-Generated Explanations (CAGE) framework for generating explanations for Commonsense Question Answering (CQA). We envision a LLM-based query rewriter to create more accurate, transparent, and controllable ranking models.

\section{Impact on Training Data and Model Training}
\label{sec:training-data}

Observational data collected from user clicks by modern search systems can serve as a valuable resource for training learning-to-rank (LTR) systems. However, it is widely recognized that implicit feedback through clicks may not necessarily indicate absolute relevance, particularly in LTR systems.
Click data is inherently biased, which can result in sub-optimal performance for ranking systems if left unaddressed. Indirect feedback can introduce several biases, including position bias~\cite{wsdm2017-joachims:position-bias}, trust bias~\cite{www2021-zheng:trust-bias}, selection bias~\cite{www2020-ovaisi:selection-bias} and exposure bias~\cite{sigir2016-wang:exposure-bias}, among others.
Biases tend to favor higher-ranked search results, which may receive more clicks from users, even if those results are not the most suitable for the users search query. 

The biases introduced by click data in search engines can have several negative consequences for both users and search engine providers. Biased search results can lead to a poor user experience, with users unable to find the information they are looking for, or being presented with irrelevant or low-quality results. Additionally, biased search results may reduce the diversity of results, reinforce stereotypes and biases, reduce competition, and erode user trust in search engines. 

We argue that our framework of collecting direct feedback data on interactive query understanding, to a large extent, reduces the over-dependence on click data and therefore substantially reduces some of the known biases.


\subsection{Direct feedback from user interaction}
\label{sec:direct-feedback}

Unlike indirect feedback, direct methods for obtaining feedback from users involve relevance-feedback approaches~\cite{relevance_feedback}, where users are asked to evaluate search results and indicate which ones are relevant to their query. 
The first consequence of such direct feedback is that it useful to refine the search query and improve the accuracy of future search results. 
Current methods for direct feedback are query autocompletions~\cite{baeza2004query}, and query suggestions~\cite{rosset2020leading, sordoni2015hierarchical}, are both compatible and complementary in our intent generation framework.
However, it should be noted that both these approaches are not (a) fully specified natural language or (b) not easily editable by the user.
To reduce query abandonment and promote user interaction in the query understanding phase, the presented machine intents must attempt to diversify over the intent space.
Diversification over the intent space in not new to IR, and has been extensively researched in the context of search rankings~\cite{xquad}.
However, we propose a research direction to obviate the need for hard rewrites. 
The user can thereafter interact and provide feedback through soft rewrites as discussed in Section~\ref{sec:framework}. 
Note that on one hand, the key challenge is to first learn a diversification strategy to avoid hard re-writes and abandonment.
On the other, provide generic \mi{}s increases the number of interactions until the final commit thereby potentially causing user fatigue.
Therefore a key challenge is to  balance specialization to capture intents close to user intent with diversification.

\mpara{RD III (Diverse Machine Intents):} \textit{How do we generate \mi{} that are diverse enough to capture possible intents for ambiguous queries? 
How do we balance diversification with specialized intents for user satisfaction?
}

Though diverse query intents are possible through a myriad of approaches ensuring that they are informative and useful for feedback is a challenge. Existing diversification methods in IR literature can be divided into \textit{implicit} and \textit{explicit} diversification methods. Implicit diversification approaches like MMR \cite{mmr} do not consider query intents and only rely on document dissimilarity. However, explicit diversification methods like xquad \cite{xquad} consider query suggestions as query aspects and ensure that atleast one relevant document for each aspect is retrieved. Potential solutions include leveraging diversification metrics to curate the training set where the query reformulations cover diverse aspects of ambiguous queries.

\subsubsection{Types of user-feedback data}
\label{sec:types-feedback-data}

User interaction with machine intents can provide a rich and nuanced source of data for improving information retrieval and natural language processing systems. There are several ways in which user interaction can inform and enrich this data. Firstly, when users agree with the machine intent that has been extracted, the user intent is described in near-complete detail. This can provide valuable information for developing and optimizing natural language understanding systems. Secondly, edited intents are a class of partial intents that represent small deviations from the actual intent. These deviations may reflect subtle differences in how users express themselves, which can provide insights into variations in language use and improve the robustness of natural language processing systems. Finally, query reformulations can serve as signals for non-relevance, and hence, clear negative feedback on the machine intents that are being presented.

The different types of training data that are generated through user interaction with machine intents can provide new research avenues for optimizing natural language processing systems. By minimizing the gap between machine and user intents, these systems can provide more accurate and relevant search results and improve the overall user experience. 


One of the immediate impacts of using \mi{} as queries is that it can eliminate the challenge of dealing with underspecified queries. This is because \mi{} can provide a more complete and specific representation of the user's information needs. This can have several benefits for improving information retrieval systems. For instance, it can make it easier to train transformer-based rankers between pairs of text that are more or less fully specified, rather than relying on underspecified queries that are prone to ambiguity and uncertainty. 
Additionally, using \mi{} as queries can make it easier for models to compute direct answers to user queries, as they can provide more accurate and specific information on what the user is searching for. 

\mpara{RD IV (Jointly training the Rewriter and the Retriever):} \textit{How do we train the rewriter and the retriever using the feedback signals so they can mutually reinforce each other?
}

As is a common practice in moder IR systems, the re-writer and ranker can be organized in a sequence and can be trained in a pipelined or jointly.
Training such models that have a purely a natural language interface between modules is new.
For rationale-based models, both pipelined~\cite{zhang:2021:wsdm:expred}, and joint training regimes~\cite{lei:2016:emnlp:rationalizing,yoon:2018:iclr:invase} have been proposed.
However, these approaches are implemented through masking the original feature space, and therefore do not induce a fully natural language interface.
However, policy gradient based reinforcement learning with human feedback approaches \cite{ppo, rlhf} could be employed  for jointly training the re-writer and retriever.
This way, not only can we ensure the framework can directly be optimized for arbitrary losses intended metric than a loss which acts as a proxy. 

\newcommand{\trec}{\textsc{Trec}}

\section{Feasibility Study}
\label{experiments}

We now present a case study to demonstrate how large language models (LLMs) can be used for developing a query re-writer.

We discuss the setup that could be used to answer the above questions. In particular, we first prompt a LLM-based re-writer in a few-shot setting using benchmark queries and their descriptions. The rewritten queries can be used, in place of the original queries, to fine-tune a ranker. We also compare the proposed approach with a fine-tuning setup where smaller generative models are fine-tuned to rewrite the query, unlike the few-shot prompting setup. It is worth noting that we do not use any explicitly provided training data for fine-tuning our re-writers. Therefore, we relied on the small amount of topic description data available in the TREC datasets. More details about the datasets, models, and experimental settings are provided to answer the above research questions.

\subsection{Datasets}

Our goal is to have natural language expansions of the query for downstream ranking. Since there are no existing datasets explicitly tackling this problem, we curate queries and their descriptions from several sources, as detailed below.

\subsubsection{Datasets for checking effectiveness of the LLM-based re-writer.}
To tackle \textbf{RDI}, we collect several datasets from \trec{} web track \cite{clarkeoverview}. Particularly, we collect topics, subtopics and corresponding descriptions from \trec{} web tracks from 2009 to 2011 to be used as dataset for fine-tuning the generative model.  
We sourced the \texttt{(topic, topic description)} pairs to train a generative model where the \texttt{topic} acts as a query and the \texttt{topic description} is supposed to the \mi{}.
We obtain 1143 samples for training, 126 samples for validation. 
Finally, we use 103 samples of the \trec{} Web 2012 topics, subtopics, and their descriptions for testing the rewriter. 

\subsubsection{Google People Also Asked Data}
Since the TREC topics/queries are under-specified and do not disambiguate the multiple intents a query could have, we propose to collect questions from the Google \textit{people also ask (PAA)} section to collect relevant questions/texts for each query. These texts from PAA could serve as query descriptions or expanded versions of the under-specified query that convey different intents. For instance, for a topic \textit{macro-molecules}, the proposed pipeline provides different questions like \textit{What are the four macro-molecules}, and \textit{What are the functions of macro-molecules?} We augment the dataset collected above with PAA questions by appending them to topics.
We fine-tune the rewriters with topics concatenated with the corresponding PAA question. However, on the test set (TREC web 2012) we leverage only the topics as access to an external knowledge base cannot be assumed during real-world deployments.


\begin{table}[ht]
\Small
\begin{tabular}{l}
\toprule 
\textbf{Prompt:}
\\
\midrule


\texttt{Query: transportation tunnel disasters} \\ Description: What disasters have occurred in tunnels used for transportation?\\ \\ 

\texttt{query: osteoporosis} \\ \texttt{description:} Find information on the effects of the dietary intakes of potassium, \\ magnesium 
and fruits and vegetables as determinants of bone mineral density in \\elderly men and women thus preventing osteoporosis (bone decay)

\\ \\

\texttt{query: Viral Hepatitis} \\ \texttt{description:} What research has been done on viral hepatitis and what \\progress has been made in its treatment?
\\ \\

\texttt{query: R\&D drug prices} \\ \texttt{description:}  Identify documents that discuss the impact of \\the cost of research and development (R\&D) on the price of drugs.
\\ \\ 

\texttt{query: dnr} \\ \texttt{description:}  [INSERT]
\\ \\ 

\bottomrule
\end{tabular}
\caption{Example of In-context learning through few-shot prompting of LLMs (limited examples shown due to space limits)}
 \label{fig:incontext}
\end{table}


We employ several generative models as the backbone for query rewriting to compare and contrast the quality of query rewrites. 

\subsubsection{\bart{}}: We employ \bart{} (base version) (140 M parameters) which has been pre-trained in a  denoising autoencoder fashion. We posit that the model could generate better natural language queries from underpsecified queries due to its function as a denoiser. We also propose a variant of the model where we augment the input topics with related paa questions through concatenation during training. We name this model \bart{} (topic+paa - target). However, both models are given only topics as input during inference. We propose the second variant to test if augmentation of paa questions to topic names during training aid in disambiguation.

We use a learning rate of 2e-5, weight decay of 0.01, batch size of 16 and train the model for 8 epochs. The same hyperparameters are adopted for all variants of the approach discussed.

\subsubsection{\gpt{}}: \gpt{} is a generative model based on the transformer decoder. We fine-tune the model to generate topic descriptions from topics on the TREC web dataset. We use the version of \gpt{} with 124M parameters.

We use a learning rate of 3.2e-5, weight decay of 0.01, batch size of 16 and train the model for 6 epochs. The same hyperparameters are adopted for all variants of the approach discussed.
    
\subsubsection{\gptt{} (prompting)}:  
\label{gpt3_prompting}
Since prompting large LMs like \gptt{} have proven to yield more stable outputs, we employ two variants of prompts to \gptt{}. We feed the prompts \textit{"Generate short sentence expanding:"} or \textit{"Generate short sentence question:"} followed by the topic name to \gptt{}. We call these two variants \gptt{} (prompt1) and \gptt{} (prompt2) respectively. We use the \textit{text-davinci-003} from Openai-API. We use a temperature of 0.5, max token length of 35 to generate short natural language rewrites of the original query. For frequency penalty and presence penalty, we use values of 0.8 and 0.6 to avoid redundancy in generated outputs.

\subsubsection{\gptt{} + in-context examples}: Since, plain prompting might result in topic drifts of generated text and the model might misinterpret the intended task, we also adopt in-context learning \cite{gpt3incontext}. In-context learning treats the LM as a black-box and instructs the model of the task without gradient descent through examples. We provide examples in form of 
$$ \mathtt{<query_1,desc_1>}, \mathtt{<query_2, desc_2>} \ldots \mathtt{query_{test}, [insert]}
$$ 
and instruct the model to fill the description in the placeholder provided. An example is shown in Figure \ref{fig:incontext}.

We use the same hyperparameters as discussed in \ref{gpt3_prompting}. Here too, we use the \textit{text-davinci-003} model from the Openai-API. We select and construct ten demonstration samples from the TREC web track \cite{clarkeoverview}. We carefully curate the examples manually for our preliminary exploration of LLM based disambiguation of query intent. We construct these samples based on common underspecified queries and their descriptions found in TREC web track, such as acronyms and other queries with limited context. Some examples are shown in Figure \ref{fig:incontext}. In the future, we plan to explore automated approaches for the selection of in-context learning examples \cite{zhang-etal-2022-active}.

\subsection{Evaluation Metrics}

To evaluate the quality of re-writes one can compute ROUGEL scores and \textsc{BERTScore} \cite{bertscore} to evaluate the rewritten queries. It is a common automated evaluation metric used in multiple text generation tasks \cite{codebertscore,bertscore_app}.
The metrics serve as a proxy to measure the ability of the generative model to expand upon the query to disambiguate various intents to aid in downstream retrieval tasks. They also measure if the queries are plausible. However, manual evaluation would give a better reflection of plausibility and faithfullness of generated rewrites.

\begin{table*}[]
\begin{tabular}{l|l|p{12cm}}

\hline
\textbf{Method }                          & \textbf{topic}      & \textbf{rewrite}                                                                                                                 \\ \hline
\multirow{3}{*}{Gpt3 in context} & EPIC study          & Find information regarding EPIC (European Prospective Investigation into Cancer and Nutrition) findings on lymphoma.             \\ \cline{2-3}
                                 & Crib death          & Find information about the causes of Sudden Infant Death Syndrome (SIDS) or crib death.                                          \\ \cline{2-3}
                                     & Define Sri         & What is the meaning of the sanskrit word sri and how is it used as a title of veneration ?                                          \\ \cline{2-3}
                                 & Alli                & Find information about Alli, an over-the-counter weight loss drug.                                                               \\ 
\cline{2-3}
                                 & hormone proteins               & Find information about the role of hormone proteins in physiology.                                                               \\
                                 \hline
\multirow{3}{*}{BART (topic + paa)}            & EPIC study & Find information on EPIC study                                                                                        \\ \cline{2-3}
                                 & Crib death          & Find information about the Crib death bed at the NIH Medical Center.                                                             \\ \cline{2-3}
                                         & Define Sri         & Find the homepage of the SRI newspaper, "The Ringer".                                          \\ \cline{2-3}
                                 & Alli                & Find information about the Alli movie, Star Trek: Voyager.                                                                       \\
        \cline{2-3}
                                 & hormone proteins                & Find information about different types of hormone proteins, including how they affect body weight, etc.\\\hline
\multirow{3}{*}{GPT (prompt1)}   & EPIC study          & The EPIC Study was a clinical trial that evaluated the safety and efficacy of a new investigational drug.                        \\ \cline{2-3}
                                 & Crib death          & Crib death, also called sudden infant death syndrome (SIDS), is the sudden, unexplained death of a baby younger than 1 year old. \\ \cline{2-3}
                 & Define Sri         & What is Standard Reference Information and how is it used to organize references ?                                          \\ \cline{2-3}
                                 & Alli                & Is Alli is a go-getter.                                                                                                        \\ 
             \cline{2-3}
                                 & hormone proteins                & Hormone proteins are proteins that are secreted by endocrine cells and act as chemical messengers \\\hline
\multirow{3}{*}{Gpt3 (prompt2)}  & EPIC study          & What is EPIC study?                                                                                                              \\ \cline{2-3}
                                 & Crib death          & What is crib death ?                                                                                                             \\ \cline{2-3}
                     & Define Sri         & What is Sri lanka ?                                          \\ \cline{2-3}
                                 & Alli                & Is alli hungry ?   \\

                   \cline{2-3}
                                 & hormone proteins                & What are hormone proteins?
                                 \\ \hline                                                                                                
          
\end{tabular}
\caption{Qualitative analysis of rewrites for underspecified queries}
  \label{tab:qualitative}

\end{table*}

\section{Qualitative Analysis}
We analyze examples of ambiguous queries and their corresponding rewrites obtained from the generative models. Some of these examples are listed in Table \ref{tab:qualitative}. We observe that the machine intents in the generated queries are close to actual intents in the \gptt{} + in-context approach. For instance, for the query \textit{EPIC study}, there are many possible expansions such as \textit{Emergency Preparedness and Injury Control}. However, the corpus only contains documents pertaining to the cancer study. This renders the query and underlying intent ambiguous without being contextualized by the documents in the corpus. However, among the other rewriting approaches \gptt{} + in-context approach is able to decipher the correct intent with fine granularity by also providing an expansion for EPIC. The resulting query is also scrutable as users can clearly decipher the machine intent based on the expansion and edit their query or the rewrite if their intent deviates from that in the rewritten query. The advantage of the envisioned system is more evident from the third example \textit{Alli}. This query is also of ambiguous nature as it could refer to a movie star or a character in movies and a weight loss drug. The other queries in the table are also of ambiguous nature. However, we observe that \gptt{} with in-context learning is able to decipher the correct intent without any additional context. We attribute this to the large scale pre-training data and the ability of over-parameterized models to serve as knowledge bases \cite{gpt3_knowledge_base}. Based on the discussed results and the success of LLM based document expansion approaches like HyDE and query2doc \cite{hyde,wang2023query2doc}, we posit that the proposed framework will further advance downstream retrieval performance. 

\textbf{Insight}. \gptt{} with in-context learning is better at generating plausible rewrites with intents that are relevant and potentially improve downstream retrieval.
%


\section{Discussion and Open Questions}
 \label{open_questions}

Our initial results suggest that not only are the \mi{} plausible, but in many cases, they are faithful to the original intent.
There is one caveat and a possible threat that should be mentioned. 
Since we are using publicly available benchmark data, there is a risk of the LLMs used in our experiments have already encountered them during their pre-training phase.
Nevertheless, this risk is slightly mitigated for the smaller models that we use and although they do not perform as well as \gptt{} variants, they are still plausible.
In the rest of the section, we discuss some of the important open questions towards making tangible progress.

\subsection{Efficiency of LLM-based rewriters}
\label{sec:efficiency}

Efficiency is a key concern and still an open question for large-scale deployment of LLMs.
Especially for large-scale applications of query rewriting we desire algorithms that are resource efficient and provide real-time searching. 
Inference issues in large language models (LLMs) can include computation time, memory usage, energy consumption, and latency. 
Current solutions for neural yet efficient ranking involve expensive training but cheaper or CPU-resident dot-products to approximate semantic similarity for rankings~\cite{leonhardt2022efficient}.
To optimize inference, various techniques such as model compression, quantization, and pruning are explored to reduce the size and complexity of LLMs without sacrificing performance. 

One research question in this area is how to distill task-specific knowledge from an LLM into a smaller model that has a smaller memory footprint and is much faster to run inference. There are several techniques that can be used for model distillation, such as knowledge distillation \cite{knowledge_distillation}, where the output of the larger model is used to train the smaller model. Another technique is to use quantization \cite{quantization,sparse_gpt}, which involves reducing the precision of the model's weights and activations to save memory and computation time.
Fast decoding \cite{fast_decoding} is another important aspect of language modeling. To generate high-quality text from these models, several decoding methods have been proposed, such as greedy search, beam search, top-k sampling, and nucleus sampling \cite{nucleus_sampling}. Greedy search is a simple method that selects the most likely next word at each step, while beam search is a more sophisticated method that keeps track of a set of the most promising partial sequences. Top-k sampling and nucleus sampling are probabilistic methods that sample the next word from the top k or from a subset of the most likely words, respectively.
Finally, Retrieval augmented language modelling could be another essential approach for fast and resource efficient query rewriting~\cite{zamani2022:reml}.

\subsection{Feedback Data and Cold Start}
\label{diversification}

One of the open problems in training the re-writers is the cold start problem in the pre-deployment stage.
In the absence of representative training data, it would be unreasonable to expect meaningful intent descriptions that are diverse enough to minimize hard rewrites. 
To ensure the generative model (rewriter) generates diverse queries, the training data must ensure high \textit{intent coverage}. 
It is important to note that for massively overparameterized LLMs careful choice of in-context learning data is more crucial than scale data. 
However, there is a direct practical trade-off between the size of the deployed model and the size of the fine-tuning data used. 
This opens up interesting avenues of future research on how to carefully select a small amount of high-quality data versus large amounts of slightly noisy data.
Our initial experiments tend to suggest that data augmentation and using externally sourced data (like PAA data) already go a long way for the existing benchmark queries.
However, approaches like dataset distillation~\cite{dataset_distillation}, active learning, and retrieval augmented data selection~\cite{khattab2022:dsp} could be used to account for carefully selecting representative data with high intent diversity.

\subsection{User Experience and personalization}
\label{sec:feedback}

The ability to understand natural language queries and provide relevant responses and actions opens up new avenues of user interaction and offers a more seamless and intuitive way of interacting with search systems. 
One open question is how to present machine intents to the user without overwhelming them. When a system offers multiple options for actions based on a single query, it can be challenging to present these options in a way that is not confusing or overwhelming to the user. Designing effective user interfaces and visualizations for presenting intents is an important research problem in this area.
Another important open question is how to make the interaction with the query-rewriter seamless. In many cases, query rewriters can provide improved query reformulations to better match a user's intent. However, the process of iteratively refining the query and refining the machine intent can be time-consuming and cumbersome. Designing more efficient and effective ways to interact with the query-rewriter is an important research problem in this area.
A related open question is how to present effective machine intents personalized to a user profile. 
User interaction data can be used to build a natural language user-profile that can be used in-context the LLM to yield effective results. On the other hand, user interaction data could be used to fine-tune a smaller language model for better reformulations. The problem of efficiently leveraging user interaction data to improve the performance of language models and query understanding systems is another important research problem in this area.

\balance
\section{Conclusion}
\label{conclusion}
In this work, we propose a framework for redefining classical IR systems. Our framework reformulates ambiguous queries to intepretable, faithful and scrutable natural language queries. A preliminary feasibility study demonstrates that generating plausible and faithful rewrites are possible. We posit that the joint training of the rewriter and ranker in our framework with direct and indirect feedback signals would yield better rewrites and ranking performance. We also propose several research directions, and challenges associated with the development of the framework.
We envision that using LLM-based rewrites would also bootstrap many subareas of information retrieval like question answering~\cite{saha2020question}, entity search \cite{meij2014entity, singh2016discovering}, temporal IR~\cite{anand2011temporal, anand2010efficient}, and medical IR~\cite{
goeuriot2016medical, luo2008medsearch}.



\end{document}